# Observation of Lattice and Andreev Bound States of Vortices in $Ba_{0.6}K_{0.4}Fe_2As_2$ Single Crystals with Scanning Tunneling Microscopy/Spectroscopy

Lei Shan*, Yong-Lei Wang*, Bing Shen*, Bin Zeng*, Yan Huang*, Ang Li†, Da Wang‡, Huan Yang*, Cong Ren*, Qiang-Hua Wang‡, Shuheng Pan†, & Hai-Hu Wen*

*National Laboratory for Superconductivity, Institute of Physics and National Laboratory for Condensed Matter Physics, Chinese Academy of Sciences, Beijing 100190, China

†Department of Physics and Texas Center for Superconductivity, University of Houston, Houston, Texas 77204-5002, USA

‡National Laboratory of Solid State Microstructures and Department of Physics, Nanjing University, Nanjing 210093, China

**For a type-II superconductor, when the applied magnetic field is higher than the lower critical value $H_{c1}$, the magnetic flux will penetrate into the superconductor and form quantized vortices, which usually are arranged in an Abrikosov lattice. For the newly discovered iron pnictide superconductors, previous measurements have shown that, in electron-doped $BaFe_2As_2$, the vortices form a highly disordered structure[1-3]. In addition, the density of states (DOS) within the vortex cores[1] do not exhibit the Andreev bound states in conventional superconductors[4-8]. In this Letter, we report the observation of a triangular vortex lattice and the Andreev bound states in hole-doped $BaFe_2As_2$ by using a low temperature scanning tunneling microscope (STM). Detailed study of the vortex cores reveals that the spectrum of the Andreev bound states inside the vortex core exhibits a distinct spatial evolution: at the center of the vortex core, it appears as a single peak at 0.5 mV below the Fermi-energy; away from the core center, it gradually evolves into two sub-peaks and they eventually fade out. The drastic differences between the vortex cores of the electron-doped and hole-doped counterparts are illusive to the pairing mechanism of the iron pnictide superconductors.**

Magnetic flux quantization is one of the important quantum phenomena in the mixed state of a type-II superconductor. The spatial distribution and the spectrum of the quasi-particle density-of-states (DOS) of the vortices, the quantized magnetic flux, can provide crucial information about the pairing mechanism. Since the first successful STM imaging of the vortex lattice and the observation of the Andreev bound states within the vortex core on a conventional superconductor 2H-NbSe$_2$[5], low temperature scanning tunneling microscopy has been used to study the vortex state in unconventional superconductors [9]. It has been demonstrated that, in cuprate superconductors, the quasi-particle DOS in a vortex core shows a suppression and two low-energy subgap peaks (or kinks) near the Fermi energy[10-14]. These results, distinct from the theoretical predictions for a superconductor with either s-wave[4-7] or d-wave[15-18] pairing symmetry, are still not well understood, but suggest that the ground state of the cuprate superconductors may be gapped, manifesting an unconventional feature of the "normal state". Recently, a new family of HTSCs, iron-based superconductors (iron pnictides), was discovered[19]. Soon after the discovery, Mazin et al.[20] suggested theoretically that the iron pnictide superconductors might have an s$\pm$ pairing, an extended s-wave pairing with opposite signs of the order parameters between the hole pockets (around Γ-Z) and the electron pockets (around M-A). Subsequent STM measurements on an electron-doped iron pnictide superconductor Ba(Fe$_{1-x}$Co$_x$)$_2$As$_2$ successfully demonstrated the observation of magnetic vortices, but they were not ordered in Abrikosov lattice[1]. More interestingly, no Andreev bound state in the vortex core has been observed, which is in striking contrast to the previous results on conventional superconductors and apparently disagrees with the current theoretical predictions for iron pnictides with either $d(x^2-y^2)$-wave and $s(x^2y^2)$-wave pairing symmetry[21]. This result has also been confirmed by another STM group[22]. Then dramatically, a recent STM measurement on another Fe-based superconductor FeTe$_{1-x}$Se$_x$ raised a flag and showed strong evidence for the s$\pm$ pairing symmetry[23]. In their article, the authors demonstrated the expected magnetic field dependence of the interference pattern due to quasi-particle scattering off magnetic vortices. Since there are many families in the iron-based superconductors, the available results of STM measurements[1,23-29] are still limited to reach a consensus. An intriguing question appears to us is how the hole-doped pnictide superconductors will behave in a magnetic field. Therefore we decide to perform the similar low temperature STM measurements on iron pnictide superconductor $Ba_{1-x}K_xFe_2As_2$, which is hole-doped from the same parent compound $BaFe_2As_2$ as for the electron-doped sample used in the previous STM experiments[1].

The single crystal $Ba_{0.6}K_{0.4}Fe_2As_2$ ($T_c$ = 37 K) samples used in our experiments were grown with self-flux method [Method-I]. They all exhibit a very clean x-ray diffraction (XRD) pattern, extremely low residual specific heat coefficient in the zero temperature limit ($\gamma_0$ = 1.24 mJ/mol-$K^2$, comparing to the normal state value $\gamma_n$>50 mJ/mol-$K^2$)[30], and a sharp superconducting transition.

In the STM experiments, the single crystal samples were cold-cleaved *in situ* of the low temperature STM chamber, then immediately inserted into the microscope, which was already at the desired temperature. Fig.1a shows a typical topographic image often obtained. Such surface is highly disordered and the tip is usually not stable enough to perform spectroscopic mapping. Sometimes we can also get another type of surface which is more stable and allows us to obtain clear topographic images and subsequently perform spatially resolved spectroscopic measurements. Fig. 1b shows an image obtained under such circumstances. In the field of view, a half unit-cell step separates a lower cleaved surface from the upper one. It can be seen that both upper and lower surfaces have a similar pattern with interlaced bright and dark domains. Such surface is very stable and can persist up to temperatures above $T_c$ as shown in Figs.1c and 1d. Since we have not been able to obtain atomically resolved images on this material, the details of the exposed surface are not clear to us at this moment. However, spatially resolved low energy tunneling spectra from the bright and dark regions show no significant difference. Fig.1e displays a line profile, in terms of height variation, obtained along the white line marked in Fig.1c. Since the constant-current topographic image signal convolutes information about the variations of both the spatial height and the integrated density of states, it is not possible to identify the origin of such domain-like surfaces merely based on the topographic images without atomic resolution.

In Fig.2b we display two sets of tunneling spectra measured along the longer white line and the short white line indicated in Fig.2a respectively. All curves exhibit a superconducting gap of ~ 4 meV determined from the two coherence peaks. The spectra are quite homogeneous within a bright domain or a dark domain, but differences between the spectra from the bright and the dark domains are noticeable. In general, the gap magnitude is a little smaller and the zero-bias conductance (ZBC) is higher in a dark region. These differences can be seen more clearly when temperature is lowered. In Fig.3, a typical spectrum from the bright region is compared in details with the one from the dark region. Besides the gap magnitude is larger, a two-gap feature is clearly visible in the spectrum taken in the bright region. In addition to the 4 meV gap, a larger gap of ~ 8 meV can be identified, while this feature is almost not visible in the spectra taken in the dark region. These two gap values, 4 meV and 8 meV, are roughly consistent with the gap values obtained in our $H_{c1}$ measurements[31]. Furthermore, many of the tunneling spectra from the bright region show negligible ZBC compared to the results obtained on the Co-doped $BaFe_2As_2$ compound[1]. The inset of Fig. 3 is a high resolution spectrum near zero bias illustrating such low ZBC in the differential conductance. Considering the measurement is performed at a finite temperature (2K), this detail can be important in identifying the pairing symmetry of the superconducitng state. It also needs to point out that not all the spectra from the bright region show zero-conductance near Fermi energy. This can probably be attributed to some low energy quasi particle excitations due to, for example impurity scattering. Thermal excitation of the finite temperature together with a gap inhomogeneity can also contribute to the finite zero-bias conductance. However, the facts of the smaller gap size and the much higher ZBC observed in the spectra from the dark region may be attributed to the microscopic differences of these two different domains.

To image vortices, we take ZBC maps simultaneously while we image the topography of the surfaces. Figure 4a shows such a conductance map recorded in a 9T magnetic field. Another conductance map recorded at 4T is shown in Fig.4c. In both maps, the vortices appear as the locus with enhanced DOS. The average flux per vortex is about $2.17\times10^{-15}$ Wb at 9T and $1.98\times10^{-15}$ Wb at 4T, consistent with the single magnetic flux quantum $\Phi_0 = 2.07\times10^{-15}$ Wb. In the simultaneously obtained topographic images shown in Figs. 4b and 4d, the vortex positions are marked as blue circles. It can be seen that the vortices arrange in a triangular pattern as for an Abrikosov lattice.

The sharp vortex image in the ZBC map indicates the significant DOS within the vortex cores. In order to reveal the local DOS variation of a single vortex, a series of *dI/dV* spectra were taken equally spaced along a line crossing the vortex as denoted by the white line in

Fig.5a. The obtained data are presented in Fig.5b. At the vortex core center, a remarkable conductance peak dominates the line shape of the spectrum. Precisely, the peak is not exactly at zero bias, instead, it locates at a finite negative bias of about -0.5 mV. This negative-bias conductance peak (NBCP) will split into two peaks when measured away from the core center, with dominant spectral weight at negative bias. Finally, the two-peak feature disappears and the spectrum evolves continuously into the one as measured at zero-field. In order to show this NBCP is not induced by possible local impurity states, we take *dI/dV* spectra along the same line at zero field and present the data in Fig.5d. It is obvious that no additional features to a clean gap structure can be seen in all these spectra, demonstrating that the observed DOS modulation is truly due to the vortex state. For highlighting this systematic evolution, in Fig.5c we show the *dI/dV* spectra taken at four typical positions. We must emphasize that this negative-bias conductance peak is a common feature of all measured vortices.

It is very intriguing that such prominent vortex core bound states have not been observed in $Ba(Fe_{1-x}Co_x)_2As_2$[1]. This discrepancy may be attributed to the stronger scattering by the in-plane Co-dopants in $Ba(Fe_{1-x}Co_x)_2As_2$. It could also signal a fundamental difference between the hole-doped and electron-doped $BaFe_2As_2$, which certainly deserves further investigation.

The existence of the pronounced vortex core bound states peaked at negative bias and the spatial evolution are also in striking contrast to the case of cuprate superconductors in which two subgap peaks were observed in vortex cores[10-14]. However, our observations of the electronic behavior of vortex cores in $Ba_{0.6}K_{0.4}Fe_2As_2$ are quite similar to what observed in the conventional superconductor 2H-NbSe$_2$[5,7]. The differences are that the bound states observed in $Ba_{0.6}K_{0.4}Fe_2As_2$ peak at a small negative bias and evolve to two peaks with dominant spectral weight in the negative one.

We have performed a numerical calculation using a 5-orbital model. The calculation reproduces the vortex core bound states and the peak-splitting. However, the asymmetry in the calculated total DOS is weak, implying possible orbital-dependent tunneling matrix element effect. A more involved theoretical calculation may need to consider the band-dependent pairing interaction.

The well ordered vortex lattice observed in $Ba_{0.6}K_{0.4}Fe_2As_2$ indicates that the disorder induced vortex pinning is weaker, compared with that in $Ba(Fe_{1-x}Co_x)_2As_2$. This is understandable since the former has off-plane doping defects, while the latter possesses an in-plane doping which generates the disorders directly on the FeAs-planes. Our results also exclude the existence of macroscopic phase separation in $Ba_{0.6}K_{0.4}Fe_2As_2$, since otherwise the vortices should be highly disordered. The observation of vortex core bound states, their negative-biased peak and the splitting in hole-doped $BaFe_2As_2$, but not in electron-doped one, will deeper the comprehension of superconductivity mechanism in iron pnictide superconductors and stimulate further investigations.

**Acknowledgements**

We thank D. H. Lee, K. Kuroki for comments and suggestions. This work was supported by the Natural Science Foundation of China, the Ministry of Science and Technology of China (973 Projects No.2006CB601000, No. 2006CB921802), and Chinese Academy of Sciences.

**Competing financial interests**

The authors declare that they have no competing financial interests.

Correspondence and requests for materials should be addressed to Lei Shan or Hai-Hu Wen at lshan@aphy.iphy.ac.cn, hhwen@aphy.iphy.ac.cn

**Figure Legends**

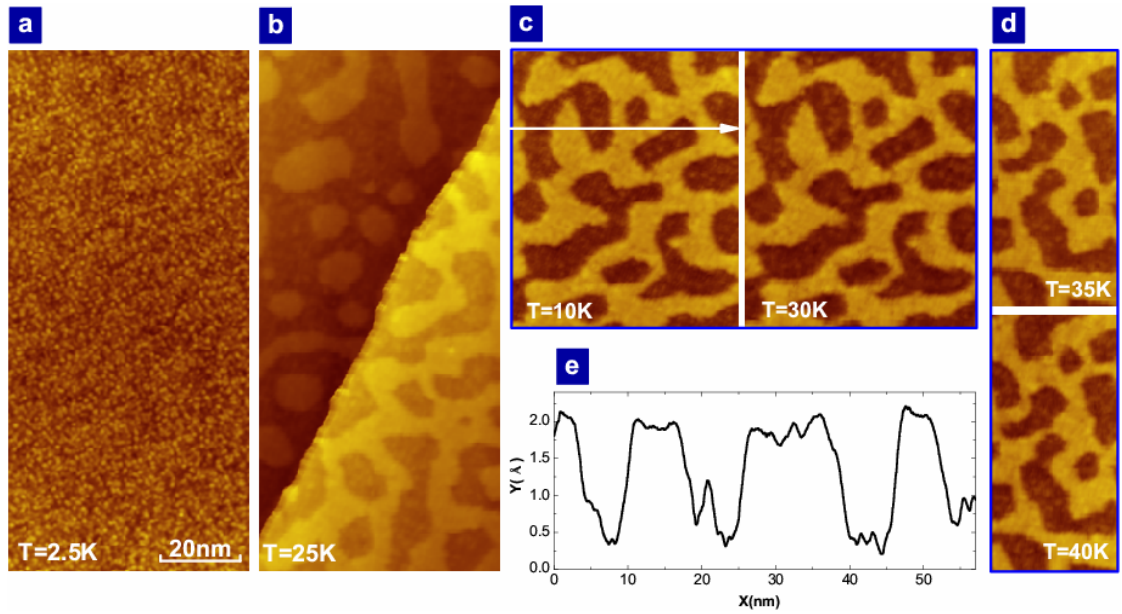

**Figure 1 Topographic STM image of the $Ba_{0.6}K_{0.4}Fe_2As_2$ single crystal cleaved *in situ*. a**, STM image taken with a sample-bias voltage $V_s$=50mV and tunneling current $I_t$=200pA. **b**, STM image of another type of cleaved surface taken with $V_s$=100mV and $I_t$=40pA, which is more stable than that shown in **a**. **c** and **d**, STM images taken on the same surface as shown in **b** while at various temperatures. No any change of the surface topography has been observed from low temperature up to above $T_c$ (37K). **e**, Section profile along the white line drawn in **c** which spans both bright and dark areas.

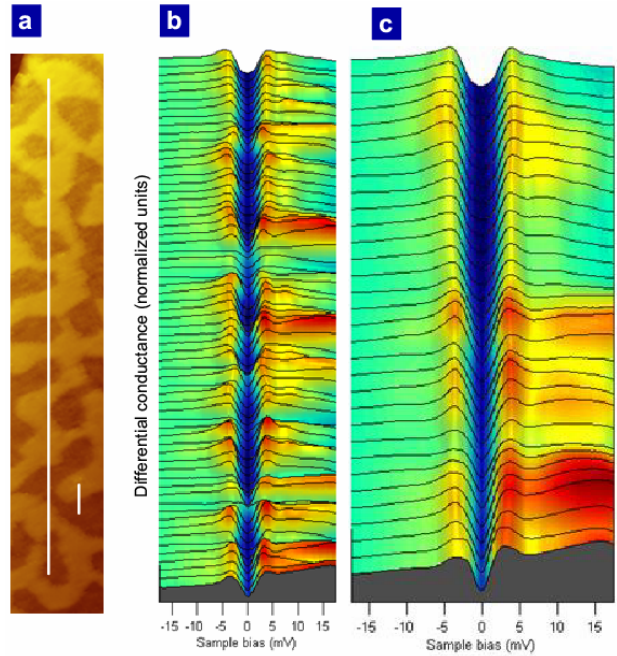

**Figure 2 Spatially resolved spectra in zero magnetic field at a temperature of 2K. a**, Topographic image of $Ba_{0.6}K_{0.4}Fe_2As_2$ taken at 2K. **b**, Differential conductance versus bias voltage (*dI/dV* spectra) measured along the trajectory with a length of 125nm drawn in **a** (the longer white line). **c**, A series of *dI/dV* spectra taken along a 10 nm line indicated by the shorter line in **a**, illustrating the variation of gap magnitude when crossing the boundary between bright and dark areas. The spectra were recorded with the settings: $V_s$=100mV, $I_t$=200pA for **b** and $V_s$=20mV, $I_t$=500pA for **c**.

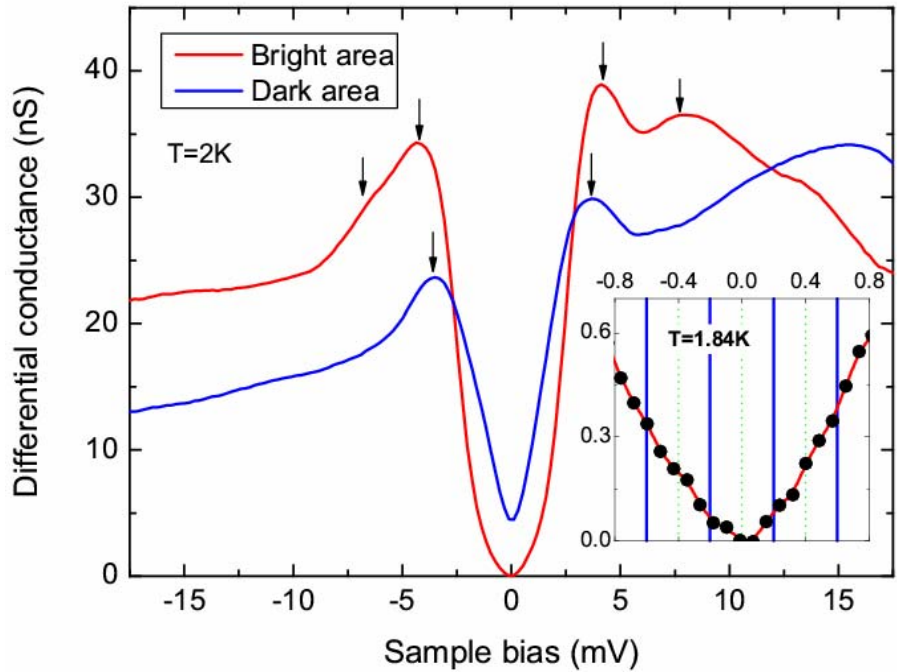

**Figure 3 Comparison of two *dI/dV* spectra measured in bright and dark areas, respectively. The superconducting gap magnitudes are estimated from the coherence peaks indicated by black arrows.** It is noted that remarkable zero-bias DOS exist in the gap of the dark-area spectrum, which is in contrast to the case of the bright-area one. Moreover, the contribution of a larger gap can be detected in the bright areas while not in the dark areas. Inset: The details of the spectrum measured at a low-bias in the bright-area.

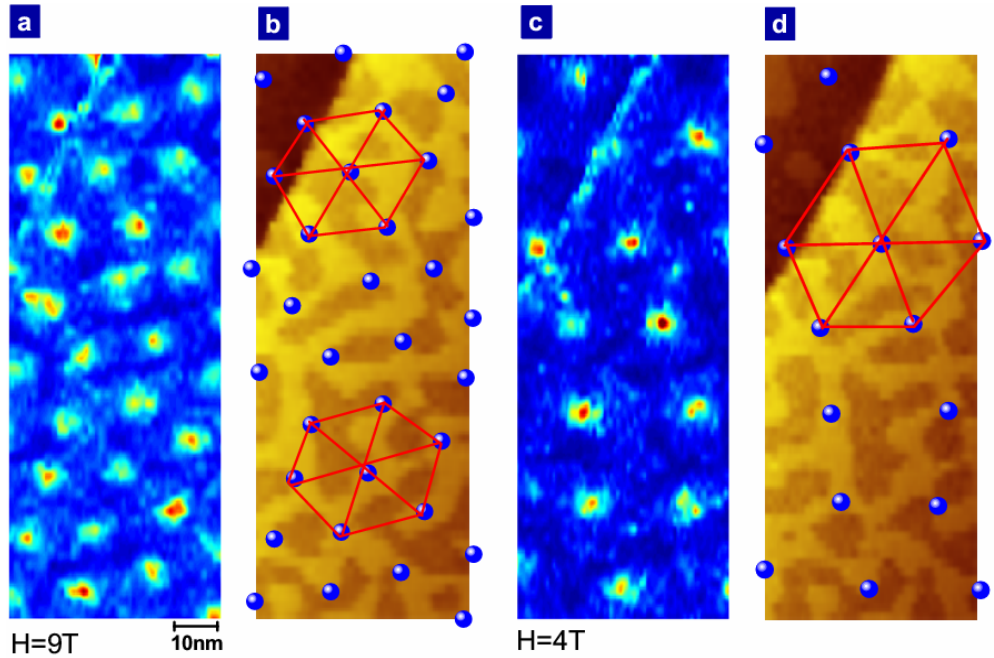

**Figure 4 Vortex lattice observed in $Ba_{0.6}K_{0.4}Fe_2As_2$ at $T$ = 2 K. a**, A 130nm×50nm zero-bias differential conductance map recorded at 9 Tesla magnetic field with the settings: $V_s$=100mV, $I_t$=200pA. **b**, Topographic image recorded simultaneously with the map in **a**, the vortex positions derived from **a** are denoted by blue circles. The triangular lattice is illustrated by red lines. **c**, Zero-bias differential conductance map measured under a magnetic field of 4 Tesla. **d**, Topographic image recorded simultaneously with the map in **c**, vortex positions and a larger triangular lattice are indicated.

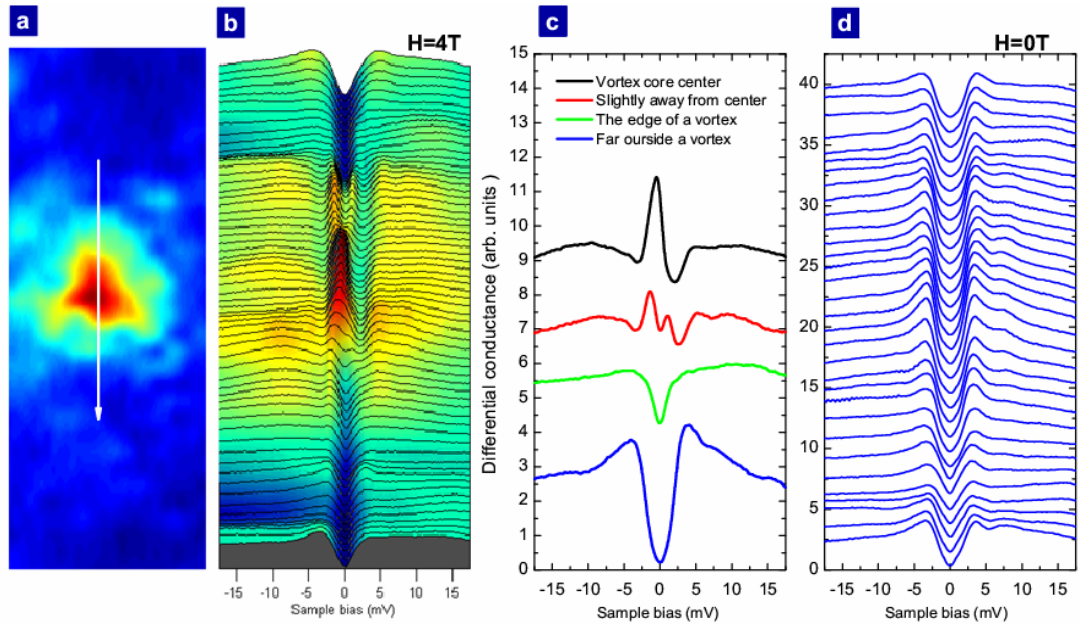

**Figure 5 Vortex core states of $Ba_{0.6}K_{0.4}Fe_2As_2$, all spectra were taken at $T$ = 2K. a**, A zero-bias differential conductance map around a single vortex measured at 4 Tesla with the settings: $V_s$=100mV, $I_t$=200pA. **b**, A series of spectra along a 104 Å trajectory (white arrows indicated in **a**) through the vortex shown in **a**. **c**, Spectra recorded at four typical positions. **d**, A series of spectra along the trajectory shown in **a** measured in zero field for comparison.

**Supplementary materials**

**Method-I: Growth and characterizations of the $Ba_{0.6}K_{0.4}Fe_2As_2$ single crystals**

The $Ba_{0.6}K_{0.4}Fe_2As_2$ single crystals were grown by using FeAs as the self-flux. The starting materials of Ba, K, and FeAs were mixed in a ratio of 0.6:1:4 and filled in a ceramic crucible. The weighing, mixing and pressing procedures were performed in a glove box filled with highly pure Ar gas, and both $O_2$ and $H_2O$ concentrations were less than 0.1 ppm. The crucible with the starting materials was sealed in a Ta crucible, which was sealed in an evacuated quartz tube. It was heated up to 1150℃, kept for 5 hours, then cooled down to 800 ℃ with a rate of 8℃/hour. The samples used in this work were 1.5mm×1.5mm×0.2mm in dimension.

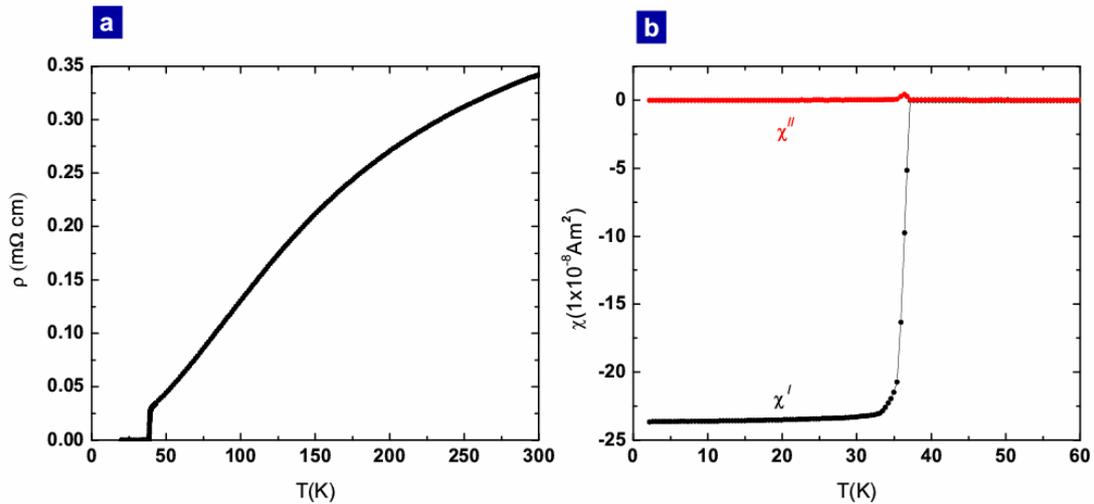

**Figure S1 (a)** Temperature dependence of resistivity of a $Ba_{0.6}K_{0.4}Fe_2As_2$ single crystal. The data were obtained with a Quantum Design Physical Property Measurement System (PPMS) with the standard four-probe method. **(b)** Temperature dependence of the ac susceptibility of the $Ba_{0.6}K_{0.4}Fe_2As_2$ single crystal. The ac susceptibility measurements were performed on an Oxford Maglab-Exa-12 system with an AC field of 0.1 Oe and the oscillation frequency of 133 Hz.

The temperature dependence of resistivity and ac susceptibility are shown in Fig.S1a and S1b. The ratio of the room-temperature resistivity over the residual resistivity just above $T_c$ is ~12, indicating that the single crystals studied here are very clean. A sharp superconducting transition can also be seen clearly in Fig.S1b with the onset temperature of 37.2K. The specific heat of the $Ba_{0.6}K_{0.4}Fe_2As_2$ single crystal was also characterized and presented in Fig.S2. The residual specific heat $\gamma_0$ obtained by extrapolating the curve $C/T$ to zero temperature is about 1.24 mJ/K$^2$·mol. Such very small residual specific heat coefficient indicates a very small nonsuperconducting volume, or weak impurity-induced quasiparticle excitations in the $Ba_{0.6}K_{0.4}Fe_2As_2$ single crystal studied here.

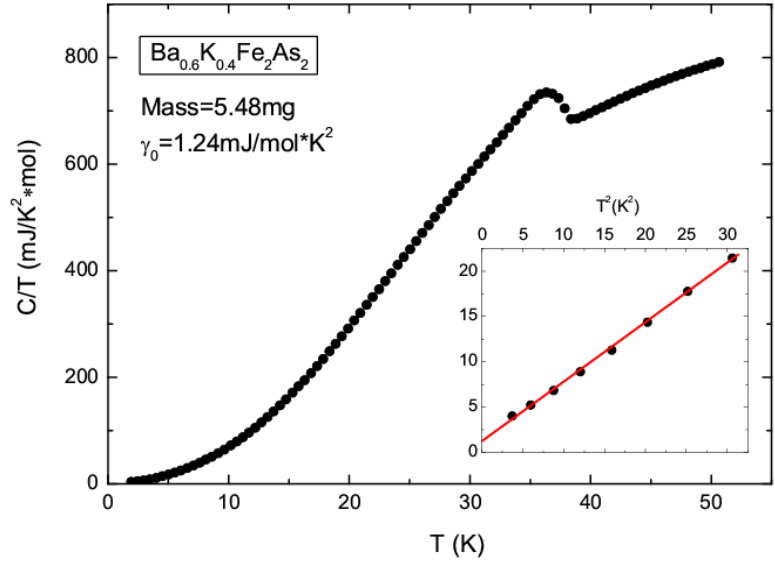

**Figure S2** Temperature dependence of specific heat of the $Ba_{0.6}K_{0.4}Fe_2As_2$ single crystal. The measurement was performed on a Quantum Design PPMS system. A specific heat anomaly appears at $T_c$. Inset: $C/T$ *vs.* $T^2$ curve at low temperatures. By extrapolating to zero temperature (indicated by the red line), the residual specific heat coefficient $\gamma_0$ can be obtained.

## Method-II: Tip preparation

All the STM/STS measurements were made using Pt-Ir (90%-10%) tips. The tips were prepared by chemical etching with saturated $CaCl_2$ solution under a higher voltage of about 30V in the beginning and then a lower voltage of about 2V. The tips were cleaned *in situ* through high-voltage (varying from several tens of voltages to several hundreds of voltages) field emission on Au in ultrahigh vacuum at low temperatures. A repeatable and very stable topographic image of Au surface can be obtained by using the prepared tips. In order to verify the good work function and flat density of states of the tips, we measured the I-V characterization between $\pm V_{max}$ with $V_{max}$ beyond 100 mV by tunneling into Au. All the tips were confirmed to have an almost linear I-V characterization before proceeding to measurements on a sample.